# Evidence of chromospheric molecular hydrogen emission in a solar flare observed by the IRIS satellite

Sargam M. Mulay,[1]⋆ Lyndsay Fletcher[1,2]†
[1]*School of Physics & Astronomy, University of Glasgow, G12 8QQ, Glasgow, UK*
[2]*Rosseland Centre for Solar Physics, University of Oslo, P.O.Box 1029 Blindern, NO-0315 Oslo, Norway*

**ABSTRACT**
We have carried out the first comprehensive investigation of enhanced line emission from molecular hydrogen, $H_2$ at 1333.79 Å, observed at flare ribbons in SOL2014-04-18T13:03. The cool $H_2$ emission is known to be fluorescently excited by Si IV 1402.77 Å UV radiation and provides a unique view of the temperature minimum region (TMR). Strong $H_2$ emission was observed when the Si IV 1402.77 Å emission was bright during the flare impulsive phase and gradual decay phase, but it dimmed during the GOES peak. $H_2$ line broadening showed non-thermal speeds in the range 7-18 km s$^{-1}$, possibly corresponding to turbulent plasma flows. Small red (blue) shifts, up to 1.8 (4.9) km s$^{-1}$ were measured. The intensity ratio of Si IV 1393.76 Å and Si IV 1402.77 Å confirmed that plasma was optically thin to Si IV (where the ratio = 2) during the impulsive phase of the flare in locations where strong $H_2$ emission was observed. In contrast, the ratio differs from optically thin value of 2 in parts of ribbons, indicating a role for opacity effects. A strong spatial and temporal correlation between $H_2$ and Si IV emission was evident supporting the notion that fluorescent excitation is responsible.

## 1 INTRODUCTION

Solar flares can have an impact on all layers of the solar atmosphere. There are strong signatures from the mid-upper chromosphere, transition region and corona, but enhanced emission corresponding to excitation of the chromospheric temperature minimum region (TMR) and possibly the photosphere is also detected. The mechanisms of excitation so deep in the atmosphere are not clear, as it seems highly unlikely that flare-accelerated electrons can penetrate there. However, depending on optical conditions, high-energy photons can. In this paper, we present observations made during a flare of enhanced line emission from molecular hydrogen, $H_2$ which has a formation temperature of 4200 K (Innes 2008) corresponding to the TMR. $H_2$ line emission is thought to be formed by photo-excitation (fluorescence) by ultraviolet (UV) radiation from the transition region, and a recent theoretical study by Jaeggli et al. (2018) put the location in a narrow range around 650 km above the photosphere for a range of temperature stratifications and radiation conditions, including corresponding to a flare. $H_2$ emission thus gives a new view of conditions in the TMR during a flare.

The formation of molecular spectra is more complex than atomic spectra. Every electronic state has multiple vibrational and rotational (sub-)states of different energies, and so excitation or de-excitation between electronic states can be between any of these vibrational or rotational states allowed by quantum-mechanical selection rules. The electronic excitation from the ground state to the first (Lyman band) or second (Werner band) electronic excited state of $H_2$ molecule occurs due to absorption of far-UV photons. Many excited vibrational levels of the upper electronic state may become populated during the electron excitation process, since there are no selection rules on the vibrational transitions. The de-excitation to the electronic ground state (with a time scale of $10^{-8}$ sec) occurs by emitting the far-UV emission lines (fluorescence) in Lyman or Werner bands of $H_2$. There are many levels in the ground state with a significant population, which gives many more options for fluorescence. More detailed information about the $H_2$ lines and their UV exciter wavelengths are given in Table 1 and in Appendix.

$H_2$ emission (in $P$ and $R$ branches that corresponds to rotational quantum number, $\Delta J$ = -1 and $\Delta J$ = +1 respectively) in solar UV spectra in the range 1175-1714 Å was first reported by Jordan et al. (1977, 1978) in observations of a sunspot umbra from the first rocket flight of the Naval Research Laboratory's High Resolution Telescope and Spectrograph (HRTS). Most of the lines observed belonged to the Lyman band of $H_2$ (Herzberg & Howe 1959; Abgrall et al. 1993a) which are fluoresced by H Lyman $\alpha$ red wing photons, and by strong transition region lines, C II, Si IV and O IV. The authors identified two groups of lines within the Lyman band of $H_2$ corresponding to two groups of transitions; the first with vibrational quantum number, $v'$ = 1 (upper level) and $v''$ = n; 2<n<9 (lower level) and the second with $v'$ = 0 and $v''$ = 4, 5 and 6.

They identified excitation by photons in the red wing of the broad and intense Lyman $\alpha$ transition region line emitted in the region of a spot or pore as responsible for the high fluorescent intensity of the $H_2$ lines. $H_2$ fluorescence in the Lyman band due to transition region lines O IV, O V, C II, C III, C IV, and Si IV was also observed

⋆ E-mail: Sargam.Mulay@glasgow.ac.uk
† E-mail: Lyndsay.Fletcher@glasgow.ac.uk

**Table 1.** Details of H$_2$ emission lines observed by IRIS in C II and Si IV windows

| Column 1 | Column 2 | Column 3 | Column 4 | Column 5 | Column 6 | Column 7 | Column 8 |
|---|---|---|---|---|---|---|---|
| H$_2$ $\lambda$ (Å) | Transition ($v'$ - $v''$) | Branch ($\Delta J = \pm 1$) | Exciting line $\lambda$ (Å) | Observed solar regions | Instruments | FWHM (Å) | References |
| 1333.475 | 0-4 | R0 | Si IV 1393.76 | Sunspot | HRTS | 0.099 | Jordan et al. (1977, 1978) |
|  |  |  |  | Flare | Skylab | – | Cohen et al. (1978) |
|  |  |  |  | Sunspot | HRTS | – | Bartoe et al. (1979) |
|  |  |  |  | Umbra, quiet region, limb | HRTS + Skylab | – | Sandlin et al. (1986) |
|  |  |  |  | Flare | IRIS | – | Li et al. (2016) |
| 1333.797 | 0-4 | R1 | Si IV 1402.77 | Sunspot | HRTS | – | Jordan et al. (1977) |
|  |  |  |  | Sunspot | HRTS | – | Bartoe et al. (1979) |
|  |  |  |  | Flare | IRIS | – | Li et al. (2016) |
| 1393.451 | 0-4 | P10 | C II 1334.53 | Sunspot | HRTS | – | Jordan et al. (1977) |
|  |  |  |  | Plage, umbra | HRTS + Skylab | – | Sandlin et al. (1986) |
| 1393.719 | 0-5 | R0 | Si IV 1393.76 | Sunspot | HRTS | – | Jordan et al. (1977) |
| 1393.732 | 1-5 | P6 | C II 1335.71 | – | – | – | – |
| 1393.961 | 0-5 | R1 | Si IV 1402.77 | Sunspot | HRTS | – | Jordan et al. (1977) |
| 1400.612 | 0-5 | R4 | O IV 1399.77 | Umbra, quiet region, limb | HRTS + Skylab | – | Sandlin et al. (1986) |
|  |  |  |  |  | – | – | Bartoe et al. (1979) |
| 1402.648 | 0-5 | P3 | Si IV 1402.77 | Umbra | HRTS | – | Jordan et al. (1977) |
|  |  |  |  |  | – | – | Bartoe et al. (1979) |
| 1403.381 | 2-6 | R2 |  | Umbra, quiet region, limb | HRTS + Skylab | – | Sandlin et al. (1986) |
| 1403.982 | 0-4 | P11 | O V 1371.29 | Light-bridge | HRTS | – | Bartoe et al. (1979) |
|  |  |  |  | Sunspot | HRTS | – | Bartoe et al. (1979) |
| 1404.750 | 0-5 | R5 | O IV 1404.81 | – | – | – | Bartoe et al. (1979) |

**Notes -** More details about IRIS H$_2$ lines can be found at https://pyoung.org/iris/

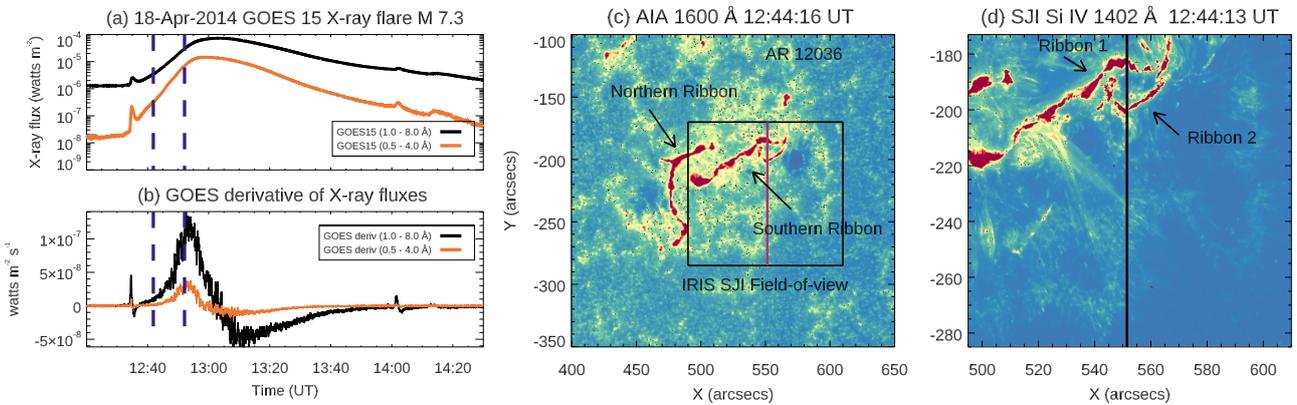

**Figure 1.** Left panel: (a) GOES X-ray M7.3 class flare observed on April 18, 2014, and (b) derivative of X-ray fluxes. The orange and black curves show X-ray fluxes in the 0.5 - 4.0 Å and 1.0 - 8.0 Å channels of the GOES-15 respectively. The blue dashed lines indicate the time slot (12:42 to 12:52 UT) during the rise phase of the flare where the physical parameters for H$_2$ line were measured. Panel (c): AIA 1600 Å image of the active region (AR) 12036. The northern and southern ribbons are shown by black arrows. The black boxed region indicate the IRIS SJI field-of-view (FOV). Panel (d): The SJI image in the Si IV 1400 Å window obtained during the rise phase of the flare. The vertical black line shows the IRIS spectrograph slit position. The slit captured emission from the southern ribbon and different parts of it are named as Ribbon 1 ('R1') and Ribbon 2 ('R2').

in a sunspot light bridge region during the second flight of HRTS (Bartoe et al. 1979).

H$_2$ emission in the Werner band (H$_2$ lines in $Q$ branch that corresponds to $\Delta J = 0$, Abgrall et al. 1993b) was found for the first time in the solar atmosphere using the first flight HRTS data (Bartoe et al. 1979). This H$_2$ emission in a sunspot (transitions in the $v'$-$v''$ = 1-5 and 1-6) was fluoresced by the O VI resonance line. The authors provided a list of wavelengths in transitions (in the $v'$-$v''$ = 1-0 to 1-7 bands) which were later found to be in good agreement with the lines observed in a sunspot (Schüehle et al. 1999) recorded by the Solar Ultraviolet Measurements of Emitted Radiation (SUMER; Wilhelm et al. 1995) instrument. H$_2$ emission has also been observed in the quiet sun (Sandlin et al. 1986) and by Innes (2008) in active region plage associated with the footpoints of X-ray microflares, near the footpoint of a brightening X-ray loop and at the location of strong transition region outflow.

Previously unidentified lines in *Skylab* flare observations by Cohen et al. (1978) were identified by Bartoe et al. (1979) as H$_2$ UV lines fluoresced by O IV, C II, C IV and Si IV. Bartoe et al. (1979) reported that these H$_2$ lines decreased rapidly in intensity with time, presumably as the line intensity and width of the exciting transition region line decreased. The spectra were recorded at the beginning of the flare gradual phase, and the spectrograph slit reportedly did not cross the flare ribbon (Bartoe et al. 1979). In contrast, the observations we report here cover impulsive and gradual phases, and the flare ribbons are in the field of view.

With data from the Interface Region Imaging Spectrograph (IRIS; De Pontieu et al. 2014), we can examine carefully the spatial and temporal evolution of H$_2$ emission, in much more detail. IRIS observes a number of molecular H$_2$ lines which mostly come from the Lyman band (Herzberg & Howe 1959; Sandlin et al. 1986). H$_2$ emission lines in IRIS spectra from the flaring chromosphere were identified by Young et al. (2015) and Li et al. (2016); H$_2$ lines are also reported in absorption, as features in Si IV spectral lines (Schmit et al. 2014). These are interpreted as due to pockets of cool (photospheric temperature) plasma, in which molecules can form, in the upper solar atmosphere above a source of Si IV emission.

Table 1 displays H$_2$ lines in the IRIS range, in the C II and both Si IV spectral windows. Column 1 indicates the wavelength of the H$_2$ lines, Column 2 specifies the transitions from lower vibrational states ($v''$) to higher vibrational states ($v'$), whereas Column 3 shows corresponding transitions in the R ($\Delta J = +1$) and P ($\Delta J = -1$) branches. Column 4 indicates the transition region lines which excite the H$_2$ emission as identified in the references given in Column 8. A possible alternative excitation route for the upper level of H$_2$ at 1333.797 Å which is analysed in this paper, involving Si IV and C II lines, is discussed in Appendix A. In Columns 5-8, we list the solar regions, instruments, and full-width-half-maximum (FWHM) of the line from the literature, if the H$_2$ line has previously been reported. A more extensive list of H$_2$ emission lines in the UV part of the spectrum between ∼ 1175 Å and 1710 Å is given by Jordan et al. (1978); Bartoe et al. (1979); Sandlin et al. (1986).

In this paper, we report on IRIS spectra of molecular hydrogen lines in the C II 1330 Å window, observed during SOL2014-04-18T13:03, an M7.3 class solar flare. The IRIS flare observation started at 12:33:38 UT, about 2.5 minutes after the start time (12:31 UT) of the GOES X-ray flare. The IRIS slit was well-positioned over the flare ribbon. The event has also been studied by Brannon et al. (2015), Brosius & Daw (2015), Cheng et al. (2015) and Brosius et al. (2016). Brannon et al. (2015) and Brosius et al. (2016) focused on coherent quasi-periodic pulsations in IRIS and EIS data, seen in both flare ribbons during the impulsive rise, finding intensity pulsations (in IRIS and EIS) and velocity pulsations (in IRIS), with different periods at different locations in the ribbon. In the IRIS observations (the same raster study we analyse here), a sawtooth pattern of Doppler shifts with an average period of ∼140 s and oscillation speed of ∼20 km s$^{-1}$ was measured using Si IV, with similar behaviour and amplitudes seen in O IV and C II lines (Brannon et al. 2015). These lines have different formation temperatures, so the similar velocity patterns led the authors to suggest that the transition region and upper chromosphere is compressed during the flare, resulting in these lines all originating from a very narrow range of heights undergoing essentially the same behaviours. Brosius et al. (2016) reported an oscillation period of 75.6±9.2 s using the EIS lines (O IV, Mg VI, Mg VII, Si VII, Fe XIV, and Fe XVI). All these lines were red-shifted, whereas a couple of components of Fe XXIII line profile were highly blue-shifted, indicating explosive chromospheric evaporation.

In this paper, we focus on the H$_2$ flare ribbon emission, produced by much cooler plasma, along with its exciting Si IV lines. The paper is organised as follows. In Section 2, we provide details about the observational study and data analysis. We discuss and summarise the results in Section 3.

## 2 OBSERVATIONS AND DATA ANALYSIS

### 2.1 Overview

The Geostationary Operational Environment Satellite (GOES-15) recorded X-ray fluxes for the M7.3 flare in two channels, 1-8 Å and 0.5-4 Å. The GOES flare[1] started at 12:31 UT, peaked at 13:03 UT and ended at 13:20 UT. The X-ray fluxes and their derivatives (which are - via the Neupert effect (Neupert 1968; Hudson 1991; Dennis & Zarro 1993) - a proxy for the hard X-ray flux) are shown in panels (a) and (b) of Fig. 1 respectively. The flare originated from NOAA active region 12036 (S15 W42[2]).

The UV signatures of the flare were observed as two bright flare ribbons located between two sunspots of configuration $\beta\gamma$. We identify the ribbons as northern ribbon and southern ribbon. These are seen in the UV images in the 1600 Å channel (see panel (c) of Fig. 1) from the Atmospheric Imaging Assembly instrument (AIA; Lemen et al. 2012) on board the Solar Dynamic Observatory (SDO; Pesnell et al. 2012), at a resolution of 0.6″per pixel and 12 sec cadence. The AIA data was obtained from the Virtual Solar Observatory[3] (VSO) and prepared using the standard AIA package `aia_prep.pro` available in the Solarsoft libraries (SSW; Freeland & Handy 1998). The black box in panel (c) of Fig. 1 shows the IRIS slit-jaw imager (SJI) field-of-view (FOV) overlaid on an AIA 1600 Å image. The entire southern ribbon and a small part of the northern ribbon were captured by the SJI.

A joint IRIS-Hinode Operation Plan (IHOP241) observational sequence was run on April 18, 2014, between 12:33 and 17:18 UT. The IRIS slit was well-positioned to cross southern ribbon and emission spectra in Si IV, O IV, C II, Mg II and H$_2$ lines were taken by the IRIS spectrograph (SG) in sit-and-stare mode, remaining stationary with respect to the solar surface at the slit position shown by the black vertical line in panel (d) of Fig. 1. Spectra were captured from the southern ribbon at around 200 slit positions. We identify different parts of this ribbon as Ribbon 1 ('R1') and Ribbon 2 ('R2'). The evolution of this ribbon was captured by the SJI (in C II 1330 Å, Si IV 1400 Å and Mg II 2796 Å windows). Table 2 summarizes the IRIS observation details.

The processed IRIS level 2 data were obtained from the archive[4]. Using the `iris_orbitvar_corr_l2s.pro` routine, the data were corrected for orbital variation (both the thermal component and the spacecraft velocity component). The dust particles on the SJI CCD produce black dots/patches in the images. They were removed using the `iris_dustbuster.pro` routine and the cosmic rays were also removed using the `despik.pro` routine. We used the strong photospheric O I 1355.6 Å line to perform an absolute wavelength calibration (see IRIS Technical Notes ITN 20).

Examples of averaged spectra at 12:51:57 UT (slit position no. 118) for both H$_2$, and both Si IV lines are shown in Fig. 2. The panels (a-c) show detector images for different IRIS spectral windows. A number of lines observed in these windows are indicated. The white dashed lines indicate the pixels along the slit which were used to obtain the example averaged spectra, shown in panels (d-f) of Fig. 2. H$_2$ at 1333.47 Å was very weak throughout the flare evolution except for slit position number 118 where the line was strong. We fitted both H$_2$ lines with a single Gaussian. The Gaussian components and the entire fitted lines are shown by blue dashed and solid red lines respectively.

---

[1] ftp://ftp.swpc.noaa.gov/pub/warehouse/2014/
[2] https://www.solarmonitor.org/?date=20140418
[3] https://sdac.virtualsolar.org/cgi/search
[4] https://iris.lmsal.com/data.html

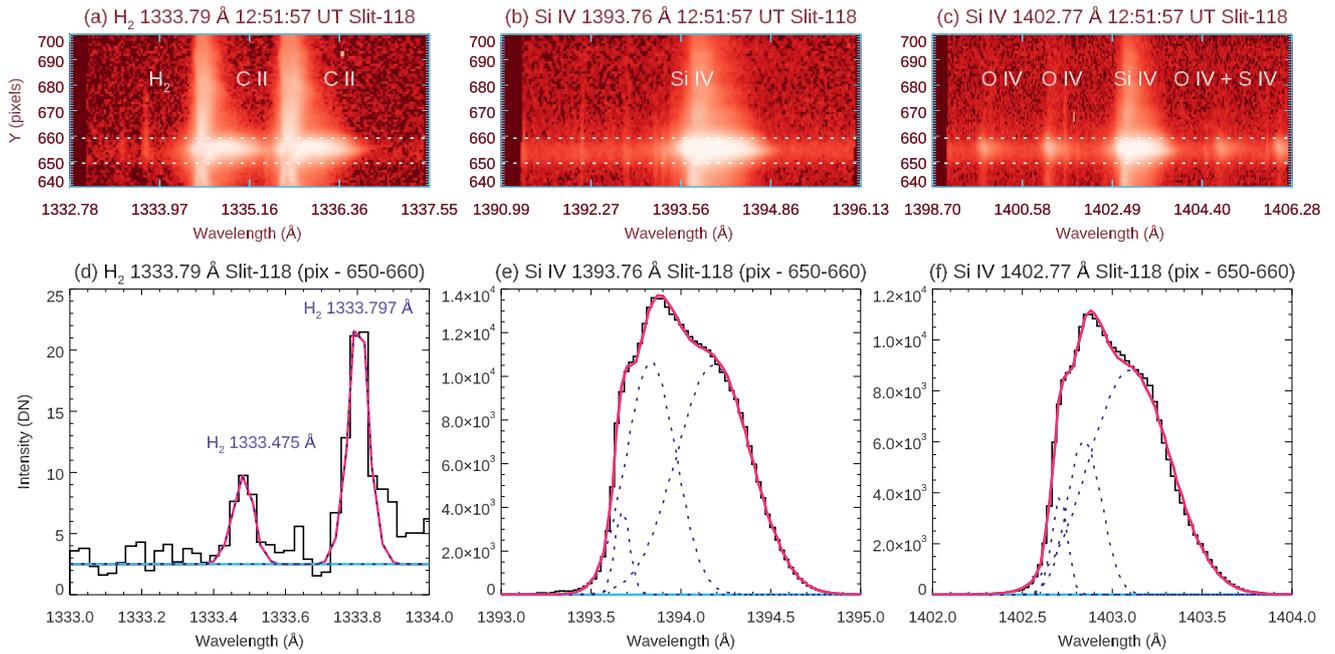

**Figure 2.** The detector images (panels (a-c)) and spectra (panels (d-f)) were obtained for slit position number 118 that shows emission at 12:51:57 UT. The spectra were obtained by averaging pixels between 650 and 660 along the slit. The white dashed lines indicate emission in these pixels. The blue dashed lines indicate Gaussian components used for fitting the lines and the entire fit is shown by solid red line. The horizontal cyan lines indicate a fit for background emission.

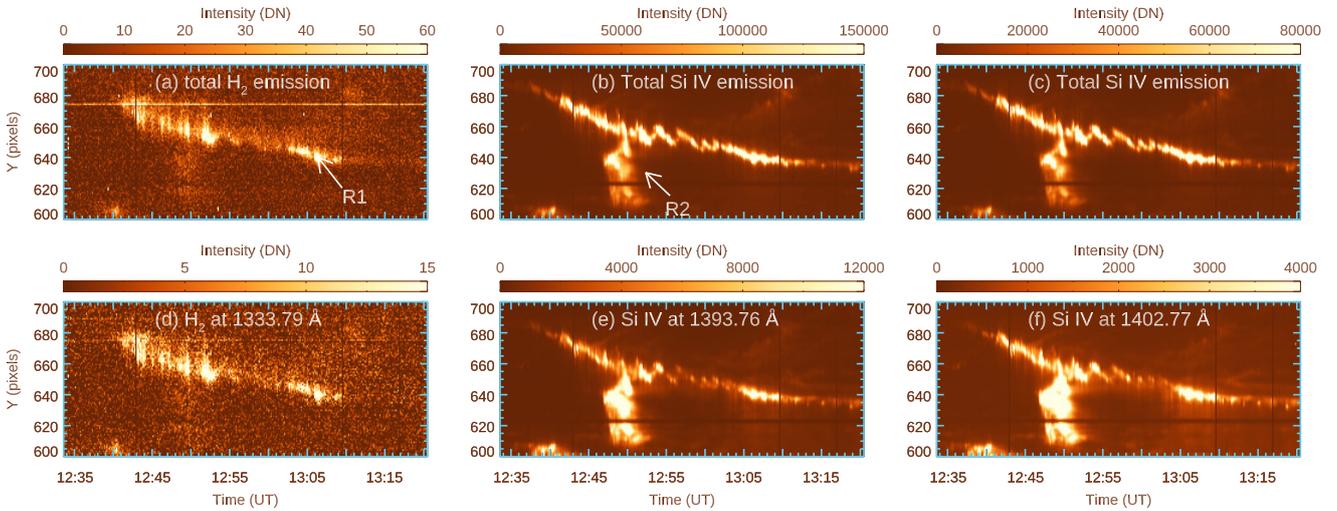

**Figure 3.** The IRIS spectral images for H$_2$ (panels (a) and (d)), Si IV 1393.8 Å (panels (b) and (e)) and Si IV 1402.77 Å (panels (c) and (f)) lines. Panel (a-c): spectral images created by summing DNs over the wavelength range 1333.76 - 1333.87 Å, 1393.5 - 1394.6 Å, 1402.5 - 1403.6 Å. R1 and R2 indicate emission from two flares ribbons (same regions that are shown in panel (d) of Fig. 1). Panel (d-f): spectral images at single wavelength values of H$_2$ at 1333.79 Å, Si IV at 1393.76 Å and Si IV at 1402.77 Å. The dark vertical lines at 12:43, 13:09 and 13:17 UT indicate that the IRIS spectra is missing at those slit locations/timings.

Based on sunspot, coronal hole, and quiet sun spectra obtained from the SUMER spectrograph, Curdt, W. et al. (2001) observed that the S I line at 1333.80 Å is very close to H$_2$ at 1333.797 Å.

In addition, Li et al. (2016) also mentioned that there is a possible blend based on IRIS flare observation. In order to identify a blend, the unblended S I line at 1401.51 Å could be taken as a reference. The behaviour (in intensity, velocity, and width) of this S I and the possibly blended H$_2$ line could be tested for evidence of correlation that would indicate an important contribution of S I to the line profile. A detailed analysis has been carried out, as discussed in Appendix B, and we observed that the behaviour of the S I 1401.51 Å line is different than that of $H_2$. Hence, we conclude that the S i line at 1333.80 Å is not blended with $H_2$ line at 1333.79 Å.

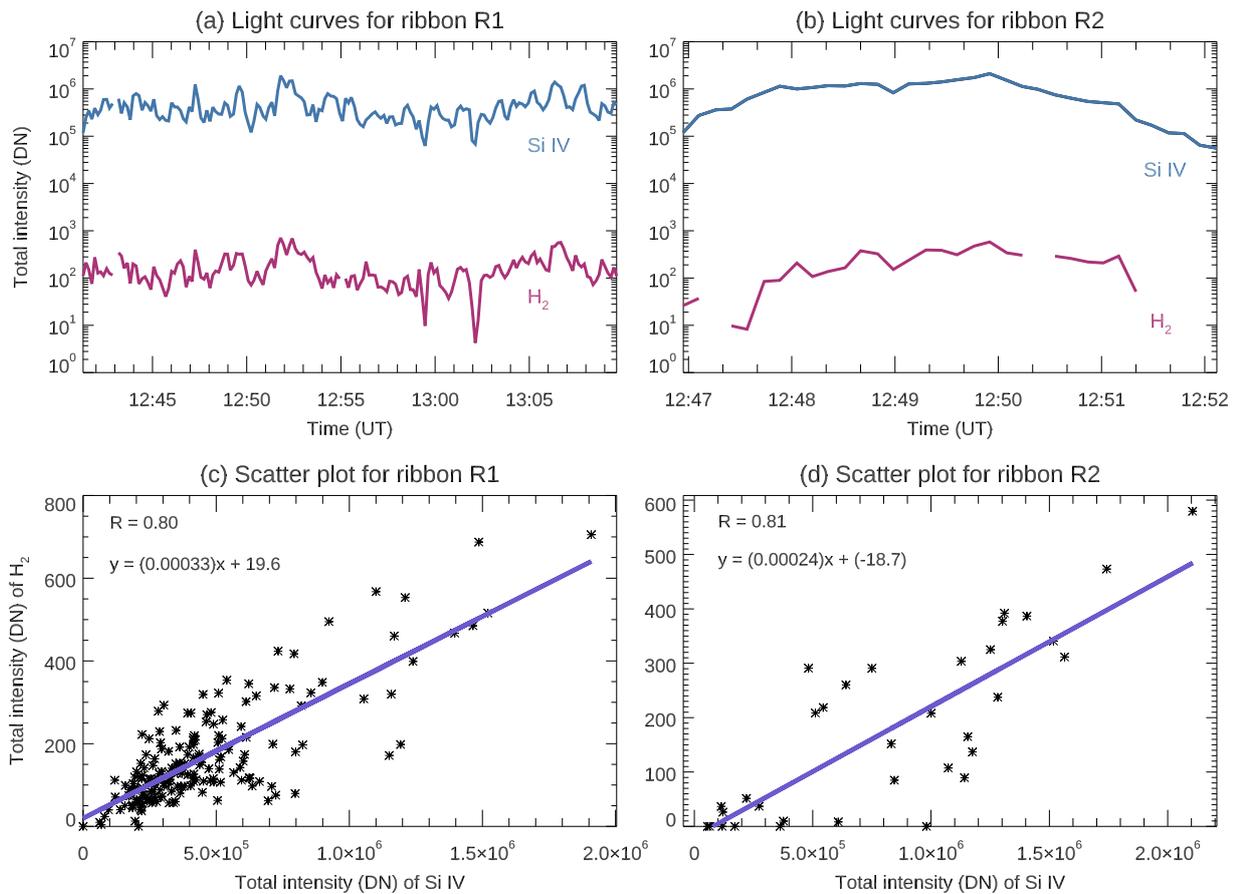

**Figure 4.** Panels (a-b): Light curves for R1 and R2. The total intensities (DNs) are obtained for the pixels where $H_2$ (between 1333.76 and 1333.87 Å) and Si iv (1402.5 and 1403.6 Å) emission is presented. The background emission (total intensity at slit 21, 12:36:46 UT, X-pix = 21 and Y-pix = 635-650) was subtracted from the total intensities before plotting. The negative intensities are removed from the data which result in discontinuities in the light curves. Panels (c-d): Scatter plots for R1 and R2 emission seen in $H_2$ (between 1333.76 and 1333.87 Å) and Si iv (1402.5 and 1403.6 Å) lines. The intensities are displayed with star symbols, and solid blue lines indicate the linear fit to the data. The equations for the fitted lines along with fit parameters are given and the Pearson correlation coefficients are displayed as 'R'.

**Table 2.** IRIS observation details of a flare

| IRIS | Spectrograph (SG) | Slit-Jaw-Imager (SJI) |
|---|---|---|
| Observation date | 18-April-2014 | |
| Observation ID | 3820259153 | |
| IHOP* | 241 | |
| Start time (UT) | 12:33:38 | |
| End time (UT) | 17:18:11 | |
| No. of rasters | 1 | 606 (No. of images) |
| No. of slit positions | 1818 | – |
| Roll angle (slit) | 0 | – |
| Step cadence (sec) | 9.4 | – |
| Spatial resolution | 0.33″ (slit width) | 0.33″ |
| Field-of-view | 0.166″ × 129″ | 121″ × 129″ |
| Exposure time (sec) | 7.9 | 1.98 |
| Cadence (sec) | 9.4 | 28 (for C ii, Si iv, Mg ii) |

*The details about the IRIS-Hinode Operation Plan (IHOP241) are available at http://www.isas.jaxa.jp/home/solar/hinode_op/hop.php?hop=0241

During the evolution of the flare, the southern ribbon was observed to move southwards in SJI images, and this displacement was nicely observed in the spectral images. The emission from R1 as it moves south is shown in the time stackplots of spectral images in Fig. 3. R1 and R2 are indicated by white arrows. R2 has a less well-defined motion. The images shown in panels (a)-(c) of Fig. 3 are created by summing DNs over the wavelength ranges 1333.76-1333.87 Å for $H_2$, and 1393.5-1394.6 Å and 1402.5-1403.6 Å for the Si iv lines.

### 2.2 Ribbon Behaviour in $H_2$ 1333.79 Å and Si iv 1402.77 Å

By selecting particular wavelengths, we can examine in some detail the ribbon evolution and correlations between the exciter wavelength and the fluorescent emission. The first thing to notice, in panels (a-c) of Fig. 3, is that the $H_2$ emission becomes visible when the Si iv 1402.77 Å becomes bright, supporting the notion that fluorescent excitation is responsible. The alternative explanation for enhanced $H_2$ emission, that the number of $H_2$ molecules has increased, is unlikely in a flare whose main outcome is chromospheric heating and thus molecular dissociation (the $H_2$ dissociation energy is 4.55 eV). In panels (d-f) of Fig. 3, we show spectral image stackplots at single wavelength values of 1333.79 Å for $H_2$, 1393.76 Å for Si iv and 1402.77 Å for Si iv lines.

These Si IV single wavelength values were chosen as they are among those responsible for exciting H$_2$ lines observed by IRIS (Table 1).

We note that if the fluorescing atom was moving at speed, so that the frequency absorbed was Doppler shifted, then the relevant exciting wavelength would have to be corrected from the values given in Table 1. However, the Doppler speeds measured for the H$_2$ lines are very small (see Section 2.3), so that the correction is much smaller than the width of one wavelength pixel, and can be ignored.

During the GOES rise phase of the flare from 12:41 to 12:55 UT, coincident with the beginning of the impulsive phase as indicated by the GOES derivative, strong emission in the H$_2$ line (summed over the wavelength range 1333.76-1333.87 Å) originated from flare ribbon R1 and very weak emission from ribbon R2 (see panel (a) of Fig. 3). H$_2$ emission from R1 reduces considerably at the GOES peak between 12:54 and 13:03 UT and then brightens again briefly between 13:04 and 13:10 UT, during the gradual phase. This is more clearly visible in panel (d) of Fig. 3, corresponding to the single wavelength pixel at 1333.79 Å. At the time of this later brightening, the GOES derivative indicates that the flare impulsive phase is over. The Si IV emission for both lines was observed throughout the flare evolution, (see panels (b) and (c) of Fig. 3) and shows many of the fine spatial and temporal details seen in H$_2$. However, there was very little emission seen at the H$_2$ fluorescent exciting frequency of 1402.77 Å between 12:55 and 13:04 UT in R1 (see panel (f) of Fig. 3) similar to the H$_2$ emission at 1333.79 Å. This is also true of Si IV at 1393.76 Å, which excites H$_2$ at 1333.475 Å (see panel (e) of Fig. 3).

Also, evident is that R2 is very bright in both the total Si IV intensity plot and the 1402.77 Å plot, but very faint in H$_2$. We examine this further by plotting light curves for ribbon R1 (between 12:42 and 13:10 UT) and R2 (between 12:47 and 12:52 UT) (see top panels of Fig. 4). At each time, the total intensities were obtained for all pixels in panel (a) of Fig. 3 where emission from H$_2$ (between 1333.76 Å and 1333.87 Å) and Si IV (between 1402.5 and 1403.6 Å) was observed. For R1, despite being almost three orders of magnitude different in DN, the overall pattern of H$_2$ intensity variation is very similar to the intensity variation in Si IV, showing many of the same small-scale features. For R2, there is little small-scale intensity variation in Si IV and H$_2$. Scatter plots for R1 and R2 (see panels (c) and (d) of Fig. 4) show a positive correlation between H$_2$ and Si IV line intensities.

As remarked above, ribbon R2 is very bright at the exciting wavelength (i.e. at Si IV 1402.77 Å) for H$_2$ 1333.79 Å, but the H$_2$ at the same time and location is faint. This is seen also in the shallower gradient for the R2 scatter plot in panel (d) of Fig. 4. To excite the H$_2$ line, the emission at 1402.77 Å must be able to penetrate to the location where molecular hydrogen is present, so it may be that the opacity of the chromosphere down to this level at the R2 location is higher than at the R1 location. It is perhaps notable that R1 crosses a plage region, whereas; R2 crosses a spot penumbra, which would be expected to have different temperature, density and hence, opacity structures.

In order to investigate this, we used a spectroscopic diagnostic tool – the intensity ratio of the resonance lines of Si IV (1393.76/1402.77) – to study the optical thickness of the plasma at the flare location. The plasma is considered to be optically thin if the Si IV intensity ratio is 2 (Mathioudakis et al. 1999). This has been used by Yan et al. (2015), who found a ratio of less than 2, with Si IV self-absorption features in transition region brightenings further indicating that opacity effects played an important role. Tripathi et al. (2020) found the ratio to be smaller than 2 at the periphery of an emerging flux region, but larger than 2 in its core. As noted, our Si IV lines exhibit complex, multi-component profiles (shown in panels (e) and (f) of Fig. 2), however if each of the components is optically thin then the ratio from the intensities integrated across the line should be equal to 2. The ratio plot is displayed in panel (a) of Fig. 5. The over-plotted black contours are Si IV emission at R1 and R2. Locations where one or both Si IV lines are saturated and the ratio cannot be evaluated are shown in the darkest red shades.

For R1, during the rising phase (12:42-12:51 UT) of the GOES flare, the intensity ratio is 2, consistent with optically thin conditions (Brannon et al. 2015), whereas; at the GOES peak between 12:52 and 13:06 UT, the intensity ratio is between 1.8 and 2.0. It is less than 2 during intervals 12:52-12:54 UT and 13:03-13:06 UT. Between 13:09 and 13:12 UT, the ratio was larger than 2 in R1, and at a substantial number of pixel locations in R2 it is larger than 2.1. However, in R2 there are also some patches between Y-pixels 628 and 650 for the interval 12:48-12:51 UT where the ratio is smaller than 1.9. The panel (b) of Fig. 5 show histograms of the Si IV intensity ratios for the two ribbons, showing a greater tendency for the ratio to be less than 2 in R1 and greater than 2 in R2. This will be discussed in Section 3, particularly in the context of flare-specific simulations by Kerr et al. (2019).

### 2.3 Spectral Analysis

We examined the spectra of H$_2$ and both Si IV lines. The H$_2$ 1333.475 Å line was very weak throughout the flare evolution except for slit position number 118 and it was difficult to fit the line with a Gaussian profile. Hence, we focused on the unblended H$_2$ line at 1333.79 Å. The counts were also low for the H$_2$ 1333.79 Å line throughout the flare evolution, so a better signal-to-noise was obtained by averaging the spectra over a number of pixels along the Y-axis (i.e. along the slit, pixel numbers are given in Column 2 of Table 3). The line was fitted with a single Gaussian profile. The IRIS instrumental width (FWHM) is about 26 mÅ (3.51 km s$^{-1}$, De Pontieu et al. 2014) for the FUV channel. The widths of the H$_2$ and Si IV line profiles were observed to be broader than both the instrumental width and the thermal width for each line, assuming that the H$_2$ emission originates from plasma at ~4200 K, and 80,000 K for Si IV. Small Doppler shifts in H$_2$ were detected. Both Si IV lines were very broad (much broader than their thermal widths), and showed non-Gaussian line profiles at each slit position (see in panels (e) and (f) of Fig. 2). Multiple Gaussian components with different red- and blue-shifts are needed to fit the Si IV lines. We do not explore this further here, but details can be found in Brannon et al. (2015) and Cheng et al. (2015). A single Gaussian component was used by Cheng et al. (2015) for simple Si IV line profiles and they measured a red shift of 12 km s$^{-1}$, and a FWHM of 26 km s$^{-1}$. Brannon et al. (2015) identified pixel positions in which Si IV lines were best fitted with two Gaussian components. They found periodically-varying Doppler velocities: the bluer component had an approximately 140 s period oscillation with a sawtooth character, varying over ±40 km s$^{-1}$, and the redder component had fluctuations around an average redshift of 50 km s$^{-1}$.

The physical parameters determined for H$_2$ 1333.79 Å emission observed at Ribbon 1 are given in Table 3. The slit position number, Y-pixel numbers along the slit, the observation time for each slit and centroids of the line are given in Columns 1, 2, 3, and 4 respectively. Columns 5, 6 and 7 indicate the FWHM of the line, nonthermal velocity and Doppler velocity of the H$_2$ line. The widths ranged from 0.037-0.084 Å, whereas; nonthermal velocities ranged from 7.1-17.8 km s$^{-1}$ and the Doppler velocities were measured for H$_2$ line. Very small red and blue shifts were obtained at ribbon R1 during the evolution of the flare, though almost all were consistent with zero within the errors.

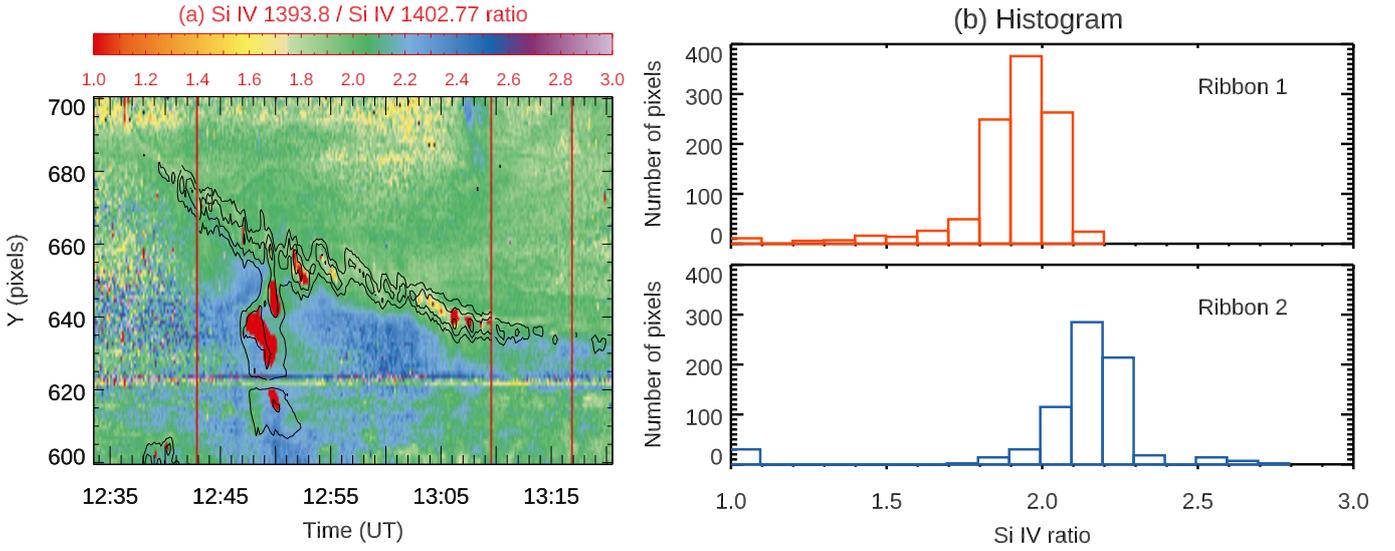

**Figure 5.** Panel (a): The intensity ratio of Si iv 1393.76 Å / Si iv 1402.77 Å is plotted. The black contours indicate emission from R1 and R2. The saturated pixels are shown in a dark red colour. Panel (b): Histograms of the Si iv line ratios for pixels in ribbons R1 and R2. The red vertical lines at 12:43, 13:09 and 13:17 UT indicate that the IRIS spectra is missing at those slit locations/timings.

**Table 3.** Parameters derived from the H$_2$ 1333.79 Å line observed at Ribbon 1

| Column 1 | Column 2 | Column 3 | Column 4 | Column 5 | Column 6 | Column 7 |
|---|---|---|---|---|---|---|
| Slit Position Number | Y-pixels along the slit | Time (UT) | Centroid (Å) | FWHM (Å) | $V_{nth}$ (km s$^{-1}$) | $V_{Doppler}$ (km s$^{-1}$) |
| 54 | 672-681 | 12:41:56 | 1333.8015 | 0.056±0.013 | 10.14±3.7 | 1.21±1.5 |
| 57 | 674-678 | 12:42:24 | 1333.7981 | 0.066±0.016 | 13.06±4.0 | 0.4±1.8 |
| 63 | 667-673 | 12:43:21 | 1333.8035 | 0.065±0.012 | 12.58±3.1 | 1.7±1.3 |
| 70 | 668-672 | 12:44:27 | 1333.8121 | 0.079±0.029 | 16.19±7.2 | 3.5±3 |
| 71 | 668-673 | 12:44:36 | 1333.8073 | 0.037±0.021 | 7.19±3.4 | 2.5±1.5 |
| 75 | 666-670 | 12:45:13 | 1333.7983 | 0.065±0.018 | 12.75±4.6 | 0.5±1.9 |
| 76 | 665-669 | 12:45:23 | 1333.8072 | 0.086±0.030 | 17.67±7.5 | 2.4±3.1 |
| 77 | 664-668 | 12:45:32 | 1333.7977 | 0.042±0.015 | 6.69±3.5 | 0.35±1.6 |
| 80 | 663-668 | 12:46:01 | 1333.7905 | 0.053±0.016 | 9.16±4.6 | -1.26±1.7 |
| 85 | 661-663 | 12:46:47 | 1333.8049 | 0.065±0.018 | 12.50±4.6 | 1.9±1.8 |
| 92 | 659-663 | 12:47:53 | 1333.7911 | 0.084±0.028 | 17.45±6.8 | -1.13±2.7 |
| 93 | 659-662 | 12:48:03 | 1333.7879 | 0.063±0.018 | 12.24±4.6 | -1.84±1.9 |
| 97 | 655-660 | 12:48:40 | 1333.8018 | 0.054±0.015 | 9.80±4.2 | 1.3±1.6 |
| 98 | 655-662 | 12:48:50 | 1333.8018 | 0.078±0.020 | 16.05±4.9 | 1.3±2.03 |
| 99 | 655-662 | 12:48:59 | 1333.8044 | 0.071±0.017 | 14.07±4.3 | 1.86±1.83 |
| 100 | 656-661 | 12:49:08 | 1333.8004 | 0.044±0.016 | 7.02±4.1 | 0.96±1.6 |
| 109 | 654-659 | 12:50:33 | 1333.8179 | 0.071±0.029 | 14.1±7.4 | 4.9±2.9 |
| 111 | 654-657 | 12:50:52 | 1333.7956 | 0.044±0.017 | 7.2±4.1 | -0.11±1.92 |
| 113 | 653-660 | 12:51:10 | 1333.8109 | 0.054±0.020 | 9.17±6.0 | 3.3±2.1 |
| 114 | 654-663 | 12:51:20 | 1333.7995 | 0.059±0.018 | 11.12±4.9 | 0.76±2.0 |
| 115 | 654-659 | 12:51:29 | 1333.8036 | 0.060±0.020 | 11.36±5.3 | 1.6±2.03 |
| 116 | 650-660 | 12:51:39 | 1333.7965 | 0.060±0.014 | 11.49±3.8 | 0.08±1.6 |
| 117 | 650-661 | 12:51:48 | 1333.7978 | 0.075±0.015 | 15.34±3.7 | 0.4±1.6 |
| 118 | 650-660 | 12:51:57 | 1333.8037 | 0.070±0.022 | 12.88±5.9 | 1.7±1.2 |

## 3 SUMMARY AND DISCUSSION

In this paper, we carried out the first comprehensive investigation of a molecular H$_2$ line observed by IRIS in a flare, revealing some properties of the cool emitting plasma. The temporal and spatial evolution of the H$_2$ emission from flare ribbons was studied and the following properties derived for this event:

- The emission in the H$_2$ line at 1333.79 Å and at its fluorescent exciting wavelength of 1402.77 Å (the Si IV line), at both ribbon locations, are strongly correlated in space and in time;
- The correlation coefficient for the H$_2$ - Si IV intensity correlations for ribbons R1 and R2 are of same order of magnitude, but the gradient is differ by ∼50%;
- The H$_2$ line is strongest during the flare impulsive phase, dims during the GOES peak, and brightens again during the gradual phase;
- The H$_2$ line is broadened, corresponding to non-thermal speeds in the range 7-18 km s$^{-1}$;
- The H$_2$ line also shows small red (blue) shifts, up to 1.8 (4.9) km s$^{-1}$
- The intensity ratio of Si IV 1393.76 Å and Si IV 1402.77 Å differs from its optically-thin value of 2 in parts of the ribbons, indicating a role for opacity effects.

Our analysis provides clear evidence that the H$_2$ line is fluorescently excited by the Si IV line at 1402.77 Å as shown by the spatial and temporal correlations observed. The Si IV line, with a formation temperature of 80,000 K, is emitted in the transition region, and some of the downwards-going radiation arrives at much deeper layers where H$_2$ is formed at a temperature of ∼4200 K. The observed Doppler and non-thermal speeds derived from the H$_2$ emission reflect the properties of this cool plasma. Measured H$_2$ Doppler shifts are consistent with zero within the errors, indicating negligible bulk flows along the line-of-sight. This is in contrast to the finding of systematic periodic blue- and red-shifts of ∼ 20 km s$^{-1}$ mean amplitude in Si IV, O IV and C II emission lines in the same event by Brannon et al. (2015). They suggested that oscillatory motion of transition region and upper chromospheric plasma on swaying field lines anchored in the deep atmosphere would give rise to the pattern of Doppler shifts. If that is the cause then one would expect the amplitude of the oscillation to decrease with increasing depth, and indeed the amplitude of this swaying motion has reduced to unobservable levels in the TMR.

There is clear evidence of non-thermal broadening of H$_2$, possibly corresponding to turbulent plasma flows, of around 10 km s$^{-1}$. Since the H$_2$ line only appears at the time of the flare we cannot say whether the flare is responsible for these non-thermal profiles, or whether they were pre-existing in the TMR and only revealed by the fluorescent emission.

The correlations coefficient between emission at 1402.77 Å and H$_2$ are of same order of magnitude in ribbons R1 and R2, but the gradient is differ by ∼50% as shown in Figure 4. This is most straightforwardly explained as being due to a lower flux of the exciter radiation arriving at the depth in the atmosphere where the H$_2$ molecule is present in R2. This could be as a result of differing optical paths between the TMR where H$_2$ is formed and the transition region where Si IV is emitted, caused by different chromospheric temperature and density structures at the two locations, through which the exciting radiation has to pass.

We cannot measure the optical depth for the *downwards-going* radiation, but can investigate the optical properties of plasma to the *outward-going* radiation, using the ratio of the two Si IV line intensities. During the impulsive phase, this ratio measured at ribbon R1 corresponds to an optically thin plasma (ratio of 2). A slight decrease in the ratio to between 1.8 and 2.0 during the peak of GOES indicates an increase in the opacity.

In the case of R2, there are many pixel locations where the ratio is larger than 2.1. It has been argued by Mathioudakis et al. (1999) that opacity effects lead to ratios below 2, with a ratio of 1.8 corresponding to an optical depth $\tau \sim 0.25$, and by Gontikakis & Vial (2018) that values greater than 2 indicate a contribution from resonant scattering of Si IV radiation. However, detailed flare radiation hydrodynamics simulations by Kerr et al. (2019) demonstrate that opacity effects can lead to a range of ratios from 1.8 to 2.3, corresponding closely to what we find, in cases where the flare energy carried by non-thermal electrons exceeds $5 \times 10^{10}$ erg cm$^{-2}$ s$^{-1}$ and/or the electron spectrum is soft. In the simulations, the opacity effects vary on timescales of seconds. The authors argue that the intensity ratio represents the ratio of line source functions, which depends strongly on the complex structure of the upper chromosphere and transition region. This requires further theoretical study.

We believe that the study of H$_2$ line and derived plasma parameters provided here will be useful for constraining further models of the chromosphere and TMR during flares.

## ORCID ID'S

Sargam M. Mulay  https://orcid.org/0000-0002-9242-2643
Lyndsay Fletcher  https://orcid.org/0000-0001-9315-7899

## ACKNOWLEDGEMENTS

SMM and LF acknowledge support from the UK Research and Innovation's Science and Technology Facilities Council under grant award numbers ST/P000533/1 and ST/T000422/1. The authors would like to thank Dr. Peter Young (NASA Goddard Space Flight Center, USA), Dr. Giulio Del Zanna, Dr. Helen Mason and Mr. Roger Dufresne (University of Cambridge, UK) and Prof. Durgesh Tripathi (Inter-University Centre of Astronomy and Astrophysics, India) for the discussion and valuable comments. IRIS is a NASA small explorer mission developed and operated by LMSAL with mission operations executed at NASA Ames Research center and major contributions to downlink communications funded by ESA and the Norwegian Space Centre. AIA data are courtesy of SDO (NASA) and the AIA consortium. NOAA Solar Region Summary data supplied courtesy of SolarMonitor.org. The GOES 15 X-ray data are produced in real time by the NOAA Space Weather Prediction Center (SWPC) and are distributed by the NOAA National Geophysical Data Center (NGDC).

## DATA AVAILABILITY

In this paper, we used the Interactive Data Language (IDL) and SolarSoftWare (SSW; Freeland & Handy 1998) packages to analyse AIA and IRIS data. All of the figures within this paper were produced using IDL colour-blind-friendly colour tables (see Wright 2017). IRIS has an open data policy. The IRIS data is available at https://iris.lmsal.com/data.html and the data[5] analysis was performed using the routines available at https://iris.lmsal.com/tutorials.html. The data and calculation of physical parameters for H$_2$ lines are available at https://github.com/SargamMulay. The AIA data is available at http://jsoc.stanford.edu/ and the data were analysed using routines available at https://www.lmsal.com/

---

[5] https://www.lmsal.com/hek/hcr?cmd=view-event&event-id=ivo%3A%2F%2Fsot.lmsal.com%2FVOEvent%23VOEvent_IRIS_20140418_123338_3820259153_2014-04-18T12%3A33%3A382014-04-18T12%3A33%3A38.xml

sdodocs/doc/dcur/SDOD0060.zip/zip/entry/. The solar flare details are obtained from the archive ftp://ftp.swpc.noaa.gov/pub/warehouse/.The GOES data analysis was performed by following IDL routines available at https://hesperia.gsfc.nasa.gov/rhessidatacenter/complementary_data/goes.html

## APPENDIX A: FORMATION OF $H_2$ LINES

The absorption of far-UV photons gives rise to electronic excitation in $H_2$. There are a number of vibrational levels in each electronic state, so de-excitation to the ground electronic state leads to the formation of $H_2$ lines at a range of wavelengths. Excitation of the upper state requires photons of specific wavelength, resulting from emission in far-UV atomic lines, or continuum, or indeed other $H_2$ molecular lines. Table A1 provides the exciting UV atomic emission lines (Column 1) for the fluorescent channels and wavelengths of interest for the IRIS $H_2$ windows we study (Column 2); the possible decay options available from the upper level (Columns 3,4); and their wavelengths/energy levels (Columns 5,6). The fluorescence lines along with their energy levels are obtained from Abgrall et al. (1993a).

If we look for example at the the $H_2$ line at 1333.797 Å, it is formed by de-excitation of a level with rotational quantum number J=1 and vibrational quantum number $v' = 0$ in the first electronic state of $H_2$. Photons at 1402.648 Å excite this upper level in an upwards transition between the ground and first electronic states. The upper excited state can then decay to any rotational and vibrational state below it that is allowed by quantum mechanical selection rules. This includes vibrational states in the electronic ground state with energies *below* that from which the upwards transition was originally excited. Therefore, the $H_2$ line wavelength can be significantly smaller than the wavelength of the photon that excited it.

In the case of the flare we examine, the UV line emission can be seen from the spectral profiles to dominate over the continuum at the fluorescent channel wavelengths. The upper levels of the $H_2$ lines are therefore excited primarily by absorption of Si IV and C II photons (and are thus a possible de-excitation option for Si IV and C II). In particular, the $H_2$ line at 1333.797 Å is excited by photons in the wing of the 1402.77 Å Si IV line.

In a theoretical study Jaeggli et al. (2018) identified a new possible wavelength at 1393.961 Å along with the previously known 1402.648 Å as a pumping source for the upper-level population of $H_2$ at 1333.797 Å. However their non-LTE modeling suggested that in flares the 1402.648 Å would be dominant.

## APPENDIX B: UNDERSTANDING A POSSIBLE BLEND OF S I (1333.80 Å) WITH THE $H_2$ (1333.797 Å)

In order to identify a possible blend of S I at 1333.80 Å with the $H_2$ at 1333.797 Å, we have taken S I 1401.51 Å line as a reference. We studied the behaviour of this line (in intensity, velocity, and width) and compared that with $H_2$ line.

We obtained 24 spectral profiles of S I 1401.51 Å line at the same slit (Column 1) and pixel (Column 2) positions as given in Table 3. Out of 24, we could fit only 11 spectral profiles with a single Gaussian component. We derived the intensities, Doppler velocities and widths of the line and compared with $H_2$ parameters. The remaining spectra were slightly narrow in the core and broader in the wings, and two Gaussian components were needed to fit the line. Hence, we did not use these line profiles for further analysis. Figure B1 shows spectral profiles of two $H_2$ lines (panel a) and the S I line (panel b) obtained at slit position number 118. The Gaussian fits for individual lines are shown.

We compared the parameters obtained from S I 1401.51 Å line profiles with those obtained from $H_2$ at 1333.79 Å. Figure B2 shows scatter plots for the intensity of lines (panel a), Doppler velocities (panel b) and widths of the lines (panel c). The Pearson correlation coefficients show weak positive correlation for the intensity and weak negative correlation for the Doppler velocities. The widths of the lines show moderate correlation. The above results confirmed that the behaviour of the S I 1401.51 Å line is different than $H_2$ line. Hence, we conclude that there is an absence of any significant S I line contribution in $H_2$ line at 1333.79 Å.

**Table A1.** Various de-excitation options for Si IV and C II lines

| Column 1 | Column 2 | Column 3 | Column 4 | Column 5 | Column 6 |
|---|---|---|---|---|---|
| Exciting line $\lambda$ (Å) | Fluorescent channel ($v'$ - $v''$) | Transition ($v'$ - $v''$) | Branch ($\Delta J = \pm 1$) | $H_2$ $\lambda$ (Å) | Wavenumber (cm$^{-1}$) |
| Si IV 1393.76 | 0-5 R0, 1393.719 | 0-4 | R0 | 1333.475 | 74992.02 |
| | | 0-5 | R0 | 1393.719 | 71750.48 |
| | | 0-4 | P2 | 1338.565 | 74706.86 |
| | | 0-5 | P2 | 1398.954 | 71481.99 |
| Si IV 1402.77 | 0-5 P3, 1402.648 | 0-4 | R1 | 1333.797 | 74973.93 |
| | | 0-4 | P3 | 1342.257 | 74501.39 |
| | | 0-5 | R1 | 1393.961 | 71738.02 |
| C II 1334.53 | 0-3 P10, 1334.501 | 0-4 | P10 | 1393.451 | 71764.30 |
| C II 1335.71 | 1-4 P6, 1335.581 | 1-5 | P6 | 1393.732 | 71749.83 |

**Notes -** The details in the Columns 1-5 are obtained from the report on molecular hydrogen by Prof. Peter Young and it is available at https://pyoung.org/iris/
Vibrational quantum number, $v'$ (upper level) and $v''$ (lower level),
Rotational quantum number, $\Delta J$ = -1 (*P* branch) and $\Delta J$ = +1 (*R* branch).
Column 6 provides a list of energy levels from Abgrall *et al.* (1993)

Based on non-LTE models, Jaeggli et al. (2018) studied the strength of $H_2$ line at 1342.256 Å in the quiet-Sun spectra. Since the line at 1333.79 Å and $H_2$ line at 1342.256 Å originate from the same upper level ($v'$ = 0), they should exhibit similar strength and properties. A discrepancy between $H_2$ (1342.256 Å) and S I (1396.113 Å) line intensities also led these authors to rule out the presence of a blend of the S I line (1333.80 Å) with $H_2$ line (1333.79 Å).

This paper has been typeset from a TEX/LATEX file prepared by the author.

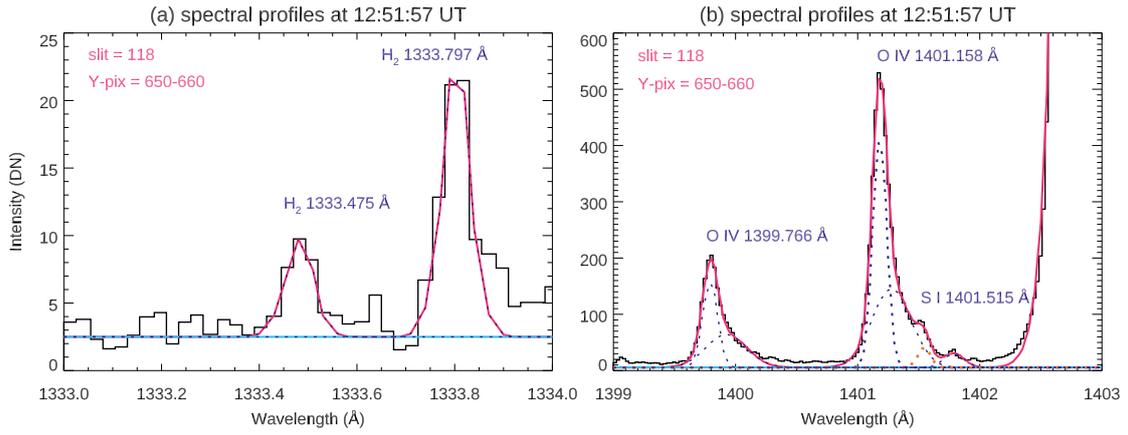

**Figure B1.** The spectral profiles of (a) $H_2$ at 1333.47 Å and 1333.79 Å and (b) S I at 1401.515 Å which are obtained for slit number 118 at 12:51:57 UT. The spectrum was obtained by averaging pixels between 650 and 660 along the slit. The blue and orange dashed lines indicate Gaussian components used for fitting the lines and the entire fit is shown by solid red line. The horizontal cyan lines indicate a fit for background emission.

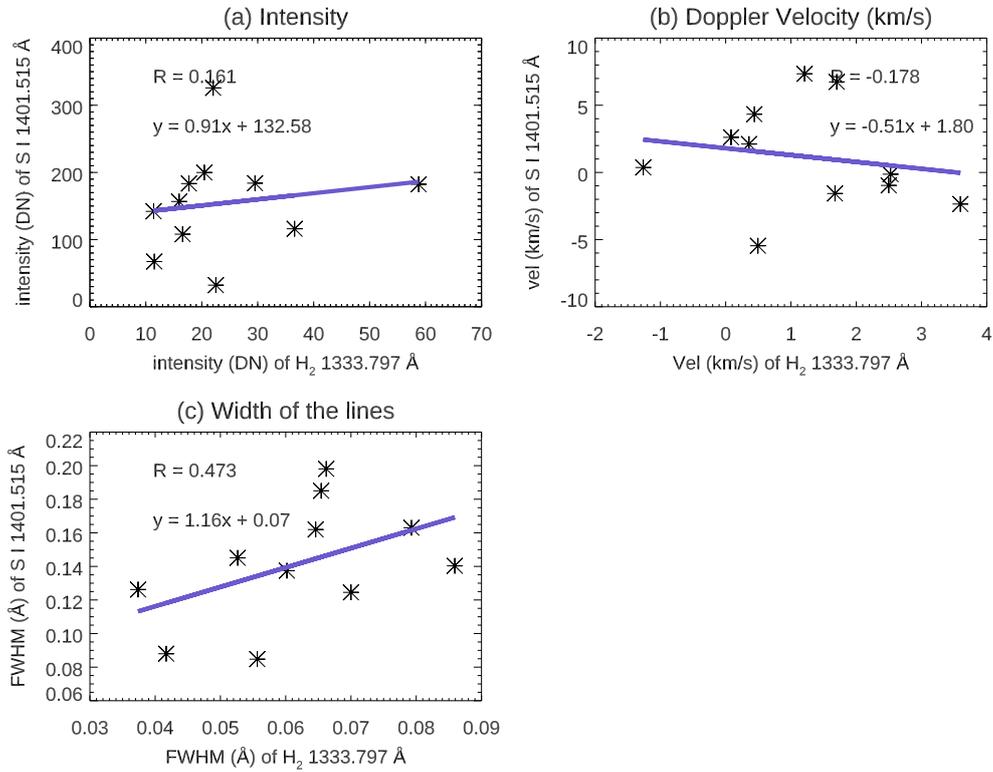

**Figure B2.** The scatter plots for the measured parameters (a) intensity of lines obtained from the single Gaussian fit, (b) Doppler velocities, and (c) width of the $H_2$ at 1333.79 Å and S I at 1401.515 Å lines. The data are displayed with star symbols, and solid blue lines indicate the linear fit to the data. The equations for the fitted lines along with fit parameters are given and the Pearson correlation coefficients are displayed as 'R'.